\documentclass[aps,twocolumn,preprintnumbers,amsmath,amssymb]{revtex4}
\usepackage{amsfonts}
\usepackage{amsmath}
\usepackage{amssymb}
\usepackage{graphicx}
\usepackage{textcomp}
\begin{document}
\title{Octave Spanning Frequency Comb on a Chip}
\author{P.~Del'Haye$^{1}$, T.~Herr$^{1}$, E.~Gavartin$^{2}$, R.~Holzwarth$^{1}$, T.~J.~Kippenberg$^{1,2}$} 
\email{tobias.kippenberg@epfl.ch}
\affiliation{$^{1}$Max-Planck-Institut f\"{u}r Quantenoptik, 85748~Garching, Germany}
\affiliation{$^{2}$\'{E}cole Polytechnique F\'{e}d\'{e}rale de Lausanne (EPFL), CH~1015, Lausanne, Switzerland}

\begin{abstract}
Optical frequency combs\citep{Holzwarth2000, Diddams2000, Udem2002} have revolutionized the field of frequency metrology within the last decade and have become enabling tools for atomic clocks\citep{Diddams2001}, gas sensing\citep{Keilmann2004, Thorpe2006} and astrophysical spectrometer calibration\citep{Murphy2007, Steinmetz2008}. The rapidly increasing number of applications has heightened interest in more compact comb generators. Optical microresonator based comb generators bear promise in this regard. Microresonator-combs\citep{Del'Haye2007} allow deriving an optical frequency comb directly from a continuous wave laser source and have been demonstrated in a number of optical microresonator geometries\citep{Del'Haye2007, Del'Haye2008, Savchenkov2008a, Grudinin2009, Agha2009, Levy2010, Razzari2010}. Critical to their future use as 'frequency markers', is however the absolute frequency stabilization of the optical comb spectrum\citep{Cundiff2003}. A powerful technique for this stabilization is self-referencing\citep{Cundiff2003, Telle1999}, which requires a spectrum that spans a full octave, i.e. a factor of two in frequency. In the case of mode locked lasers, overcoming the limited bandwidth has become possible only with the advent of photonic crystal fibres for supercontinuum generation\citep{Ranka2000, Hansch2006}. Here, we report for the first time the generation of an octave-spanning frequency comb directly from a toroidal microresonator on a silicon chip. The comb spectrum covers the wavelength range from $990$~nm to $2170$~nm and is retrieved from a continuous wave laser interacting with the modes of an ultra~high~Q microresonator, without relying on external broadening. Full tunability of the generated frequency comb over a bandwidth exceeding an entire free spectral range is demonstrated. This allows positioning of a frequency comb mode to any desired frequency within the comb bandwidth. The ability to derive octave spanning spectra from microresonator comb generators represents a key step towards achieving a radio-frequency to optical link on a chip, which could unify the fields of metrology with micro- and nano-photonics and enable entirely new devices that bring frequency metrology into a chip scale setting for compact applications such as space based optical clocks.
\end{abstract}
\maketitle

\begin{figure*}[pbth]
\begin{center}
\includegraphics[width=0.8\textwidth]{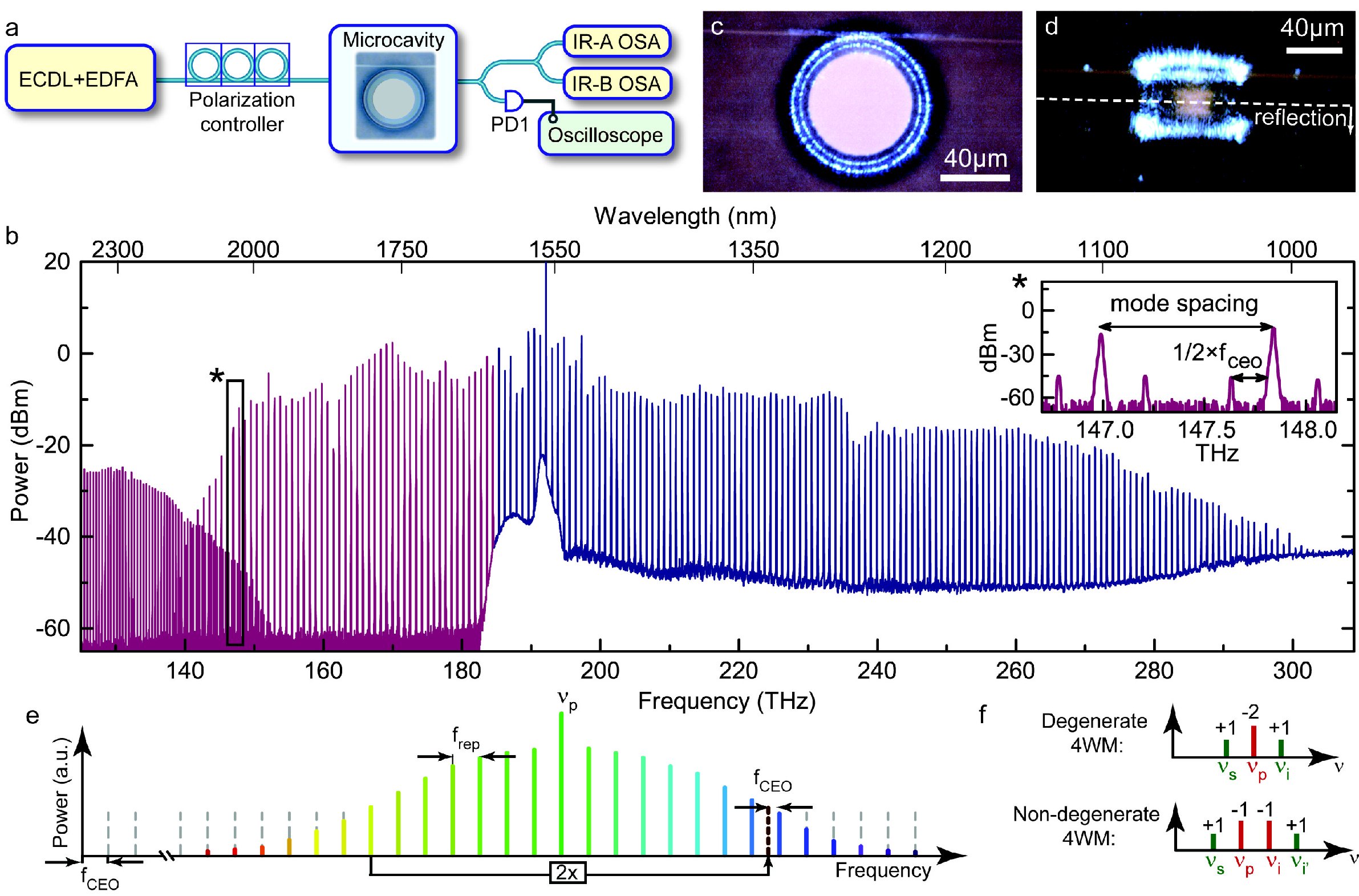}
\end{center}
\caption{\textbf{Octave spanning frequency comb generation in a microresonator.} \textbf{a.} Experimental setup. ECDL=external cavity diode laser, EDFA=erbium doped fibre amplifier, PD=photo diode. \textbf{b.} Optical frequency comb, directly generated from a cw-laser in an $80$-$\mu$m-diameter microresonator. The denser lines on the low frequency side are a replica of the high frequency end of the spectrum at half the frequency due to the grating spectrometers second order diffraction. This artefact allows determining the approximate magnitude of the carrier envelope offset frequency directly from the optical spectrum (inset). \textbf{c,d.} Microscope images of the fused silica microresonator. The comb extends into the sensitive wavelength range of the camera's CMOS sensor and can be seen as white-bluish light. \textbf{e.} Principle of f-2f-interferometry for measurement of the carrier envelope offset frequency. \textbf{f.} Microresonator-based frequency comb generation via degenerate and nondegenerate four-wave mixing (4WM). The numbers indicate the number of photons added/subtracted to the respective mode in each 4WM-process.}
\label{fig1}
\end{figure*}

In addition to the advantage of compact integration, microresonator based frequency combs have high power per comb line, which is a result of the smaller resonator size and the correspondingly higher repetition rate. This high power per comb line enables direct comb spectroscopy\citep{Diddams2007} and is advantageous for many applications, and critical for high capacity telecommunication. On the other hand, the high repetition rate of micro-combs results in a smaller peak power of the optical pulses that are underlying the generated frequency comb. This renders spectral broadening using nonlinear fibres\citep{Ranka2000, Holzwarth2000} inefficient. Spectral broadening is a method which allowed for the first broadening of mode locked lasers to octave spanning combs ten years ago\citep{Hansch2006}, leading to a breakthrough of the optical frequency comb technology.

Here we show that the high power enhancement in a microresonator itself is sufficient for direct octave spanning frequency comb synthesis without the need for any additional spectral broadening. Optical frequency comb generation in microresonators results from the interaction of the intense light field with the optical modes of the resonator via the Kerr nonlinearity (i.e. the intensity dependent refractive index). In addition to fused silica microtoroids\citep{Del'Haye2007, Del'Haye2008}, this method of comb generation has already been demonstrated in a variety of systems, including crystalline calcium fluoride resonators\citep{Savchenkov2008a, Grudinin2009}, silicon nitride microresonators\citep{Levy2010} as well as compact fibre cavities\citep{Braje2009} and silica glass ring on chip microresonators\citep{Razzari2010}. In a first step, the continuous wave light of a pump laser at frequency $\nu_{\mathrm{P}}$ is converted into signal~$\nu_{\mathrm{S}}$ and idler~$\nu_{\mathrm{I}}$ sidebands via degenerate four-wave mixing\citep{Stolen1982, Kippenberg2004a, Savchenkov2004, Del'Haye2007} (cf. Figure~\ref{fig1}f). Due to energy conservation these sidebands are equidistant with respect to the pump frequency ($2\nu_{\mathrm{P}} \to \nu_{\mathrm{S}}+\nu_{\mathrm{I}} $) and the spacing is determined by the free spectral range of the microresonator (given by the inverse cavity round trip time). The second step and actual comb generation is a result of nondegenerate four-wave mixing in which the spacing of the existing comb modes is transferred to higher order sidebands (e.g. $\nu_{\mathrm{P}}+\nu_{\mathrm{S}} \to \nu_{\mathrm{I}}+\nu_{\mathrm{S}^{\prime}}$ or $\nu_{\mathrm{P}}+\nu_{\mathrm{I}} \to \nu_{\mathrm{S}}+\nu_{\mathrm{I}^{\prime}}$).

At higher pump powers this process leads to a mixing between many different comb lines, resulting in a massive cascade of optical modes. A similar comb generation principle via non-degenerate four-wave mixing without a resonator has been recently shown using two high power seed lasers interacting in a highly nonlinear fibre\citep{Cruz2008, Cruz2009}. An optical frequency comb has two degrees of freedom, the spacing between the comb lines (repetition rate) $f_{\mathrm{rep}}$ and the carrier envelope offset frequency $f_{\mathrm{ceo}}$, such that the $n^{\mathrm{th}}$ comb line is defined as $\nu_{\mathrm{n}}=f_{\mathrm{ceo}}+ n \cdot f_{\mathrm{rep}}$.

Pivotal to the use of frequency combs in frequency metrology is the absolute control and stabilization of these two parameters ($f_{\mathrm{ceo}}$, $f_{\mathrm{rep}}$). A powerful and widely adopted method is self-referencing using an f-2f interferometer. In the latter the carrier envelope frequency $f_{\mathrm{ceo}}$ is determined by recording a beat note between the frequency doubled low frequency part of the comb and the high frequency part of the comb $f_{\mathrm{ceo}}=2 \cdot \left(f_{\mathrm{ceo}} + n \cdot f_{\mathrm{rep}}  \right)-\left(f_{\mathrm{ceo}}+ 2n \cdot f_{\mathrm{rep}} \right)$.
Implicit to this scheme is an optical comb spectrum that spans at least an octave in frequency. While it has been shown that both repetition rate and carrier envelope frequency can be stabilized and controlled\citep{Del'Haye2008}, stabilization using self-referencing has so far never been demonstrated for microresonator based frequency combs due to the lack of an octave spanning spectrum.

The experimental setup for octave spanning frequency comb generation in a toroidal microresonator is depicted in Figure~\ref{fig1}a and consists of an external cavity diode laser (ECDL) that is amplified by an erbium doped fibre amplifier (EDFA) to an output power of up to $2.5$~W. The tuning range of the diode laser extends from $1550$~nm to $1620$~nm, covering most of
the gain bandwidth of the amplifier. The amplified light is coupled into a high finesse microresonator mode using a tapered optical fibre\citep{Knight1997}. The coupling to the microresonator mode can be optimized by adjusting polarization, distance between tapered fibre and resonator, and fibre diameter. This optimization is done while periodically sweeping the diode laser output frequency across a microresonator mode of interest and monitoring the transmission signal through the tapered fibre with a fast photodiode and an oscilloscope.
Owing to the high power coupled into the resonator, the optical modes experience a significant thermal frequency shift as a result of light absorption in the microresonator. This shift can be as high as $1$~THz and a corresponding sweeping range of the diode laser is required to optimize the coupling. The employed ECDL is swept by means of a stepper motor that controls the cavity size and allows for periodic sweeping across a $1$~THz range at a frequency of $2$~Hz. Once the coupling between tapered fibre and microresonator is optimized, the laser is manually tuned into a resonance from high to low frequencies, which allows for thermal locking\citep{Carmon2004} of the microresonator mode to the pump laser. The optical frequency
comb generated in the microresonator is monitored by two optical spectrum analyzers (OSA) covering a wavelength range from 0.6~$\mu$m to 1.7~$\mu$m (IR-A OSA) and from 1.2~$\mu$m to 2.4~$\mu$m
(IR-B OSA), respectively.

Figure~\ref{fig1}b shows a measured frequency comb spectrum, spanning more than one octave (from $990$~nm to $2170$~nm), directly generated in a monolithic microresonator at a pump power of $2.5$~W and wavelength of $1560$~nm. The spectrum has been recorded using a microresonator with a mode spacing of $850$~GHz, corresponding to a diameter of $80$~$\mu$m. The more narrowly spaced modes on the low frequency side of the spectrum are artefacts from the IR-B spectrum analyzer and are actually located at twice the frequency, coinciding with the high frequency ($1$~$\mu$m) end of the comb. This artefact is generic to a grating based spectrum analyzer, as the second order diffraction of the $1$-$\mu$m light coincides with the first order diffraction of the $2$-$\mu$m-comb modes. However, these artefacts have the useful benefit of allowing direct determination of the value of the carrier envelope offset frequency from the optical spectrum analyzer (within the uncertainty of the grating based analyzer), which corresponds to twice the distance between one of the artificial second order diffraction lines (exhibiting the frequencies of the high frequency end of the comb divided by two) and an actual comb line
(cf. inset Figure~\ref{fig1}b).

It is interesting to analyze the octave spanning comb for its spectral properties. Optical frequency combs generated via four-wave mixing obey momentum and energy conservation as well as {\it photon number} conservation. The latter two conserved quantities can be interpreted as a conservation of the 'centre of mass' of the measured comb spectra, which mathematically corresponds to the first moment of the normalized photon number distribution and constantly stays at the pump laser frequency  $\nu_{\mathrm{P}}$, 
\begin{equation*}
\sum\limits_{\mathrm{i}}\nu_{\mathrm{i}} \cdot \frac{n_{\mathrm{i}}}{N}=\nu_{\mathrm{P}}
\end{equation*}

, where  $\nu_{\mathrm{i}}$ and $n_{\mathrm{i}}$ are the frequencies and photon numbers in the  i$^{\mathrm{th}}$ comb mode, and $N$ is the
total number of photons in the cavity ($N=\sum\limits_{\mathrm{i}}n_{\mathrm{i}}$). 

This conservation of the 'centre of mass' is reflected in the envelope of the frequency comb in Figure~\ref{fig1}a, showing an asymmetric comb around the pump laser, with higher power per comb mode but less lines on the low frequency side with respect to the pump laser. The 'centre of mass' of the measured spectrum in Figure~\ref{fig1}a is shifted by only $220$~GHz from the pump laser towards lower frequencies, indicating that to a very good approximation the comb is generated in a pure four-wave mixing process, even though wavelength dependent losses, Raman scattering (which would amplify the red portion of the spectrum) and coupling to the tapered optical fibre have not been taken into account.

\begin{figure*}[ptbh]
\begin{center}
\includegraphics[width=0.8\textwidth]{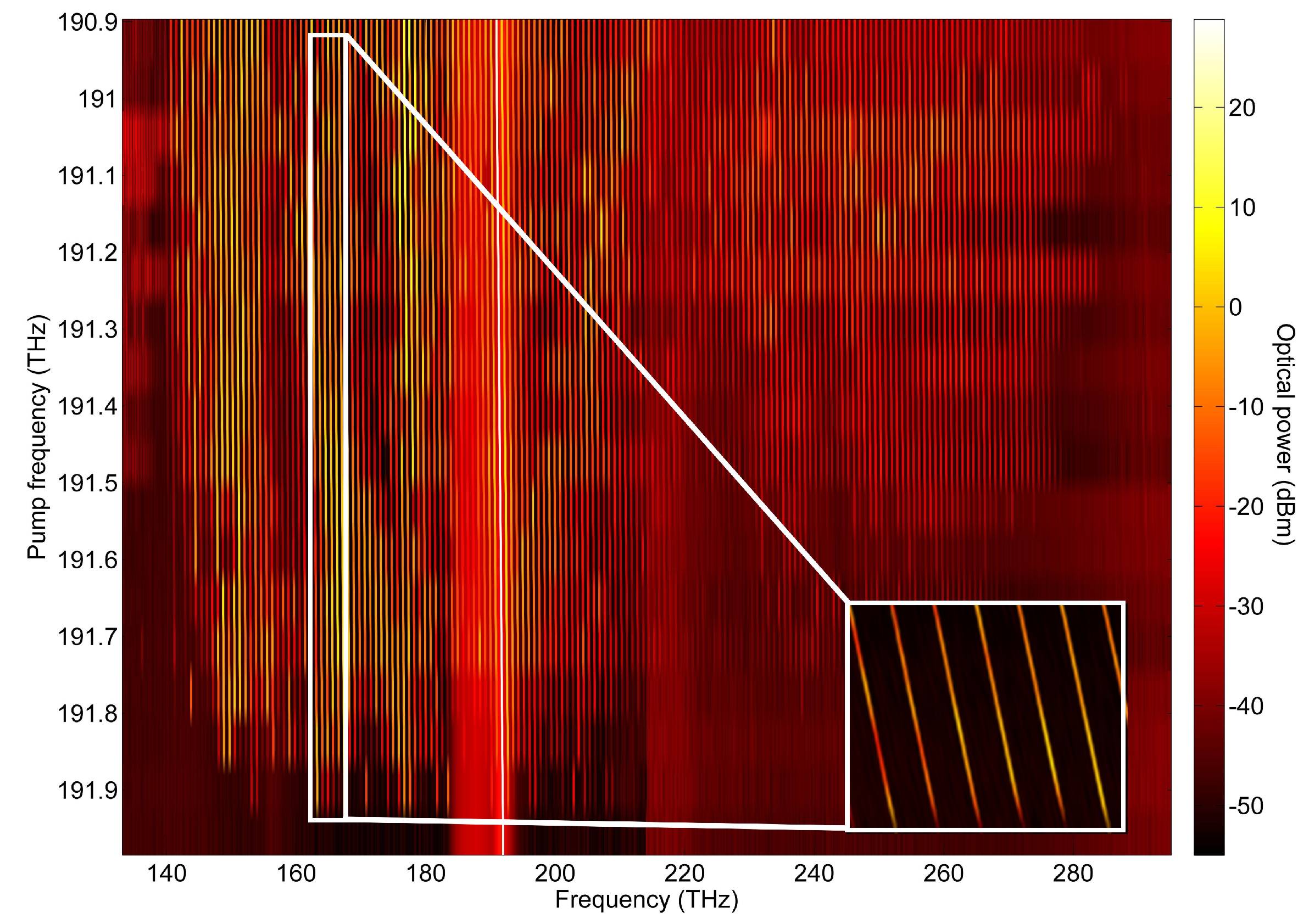}
\end{center}
\caption{\textbf{Tunable octave spanning microresonator based frequency comb.} The horizontal axis shows the measured frequency comb at different pump laser frequencies (vertical axis). The brightest line corresponds to the pump laser (with a power of $1$~W in this measurement). It can be seen that the whole frequency comb is uniformly shifted by more than one mode spacing as the pump frequency is detuned (inset).}
\label{fig2}
\end{figure*}
 
The photographs in Figure~\ref{fig1}c and Figure~\ref{fig1}d show top and side view of a microtoroid coupled to a tapered optical fibre. The white-bluish light emission from the toroid originates from the comb lines below $\sim 1.1$~$\mathrm{\mu m}$, which can be detected by the CMOS sensor of the camera.

An important prerequisite for many applications of an optical frequency comb is its tunability or in other words the possibility to precisely control the position of the comb lines. Especially for spectroscopy with a frequency comb with large mode separation, it is crucial to be able to precisely position a comb mode at any frequency of interest. While basic control and stabilization of both the mode spacing and the offset frequency of a microresonator based frequency comb have been shown in previous work\citep{Del'Haye2008}, here we present full tunability of the pump laser frequency (which defines one of the comb lines) {\it over more than an entire free spectral range} of the microresonator.

This remarkable tuning capability is shown in Figure~\ref{fig2}, depicting the colour coded optical comb spectrum at different pump frequencies (vertical axis). The thermal effect allows the microresonator modes to follow the pump laser frequency
in a tuning range exceeding $1$~THz, while an octave spanning frequency comb from $140$~THz to $280$~THz is maintained over nearly $500$~GHz tuning range. Continuous tuning of $f_{\mathrm{ceo}}$ via pump frequency tuning is a distinguishing feature of the microresonator based combs and has so far not been achieved with optical frequency combs based on mode locked lasers. The inset of Figure~\ref{fig2} is a zoom into a part of the spectrum, showing a shift of the comb modes of more than one free spectral range of the microresonator. The microresonator resonance pumped by the ECDL is shifted from $1561.8$~nm to $1570.2$~nm, equating to a continuous frequency shift of $1$~THz. Since this frequency shift only depends on material properties of the microresonator, namely the thermal expansion, thermally induced refractive index change and the optical Kerr effect, it is possible to derive the temperature change of the resonator material. For this calculation the contribution of the Kerr effect is subtracted first, being responsible for a substantial resonance shift of $\Delta \nu_{\mathrm{Kerr}} \approx 40$~GHz, which can be derived from equation~\ref{KerrShift} using the parameters for a $80$-$\mathrm{\mu m}$-diameter microresonator with a measured optical quality factor of $Q \approx 2 \times 10^{8}$, a mode cross section of $A_{\mathrm{eff}} \approx 4 \mathrm{\mu m}^{2}$ and a launched power of $P_{\mathrm{in}} \approx 1$~W:

\begin{equation}
\label{KerrShift}
\Delta \nu_{\mathrm{Kerr}}\approx\frac{n_{\mathrm{2}}}{n^{2}_{0}} \cdot \frac{c}{4 \pi^2 \cdot R \cdot A_{\mathrm{eff}}} \cdot \frac{P_{\mathrm{in}}}{Q}
\end{equation}

\begin{figure}[ptbh]
\begin{center}
\includegraphics[width=1\linewidth]{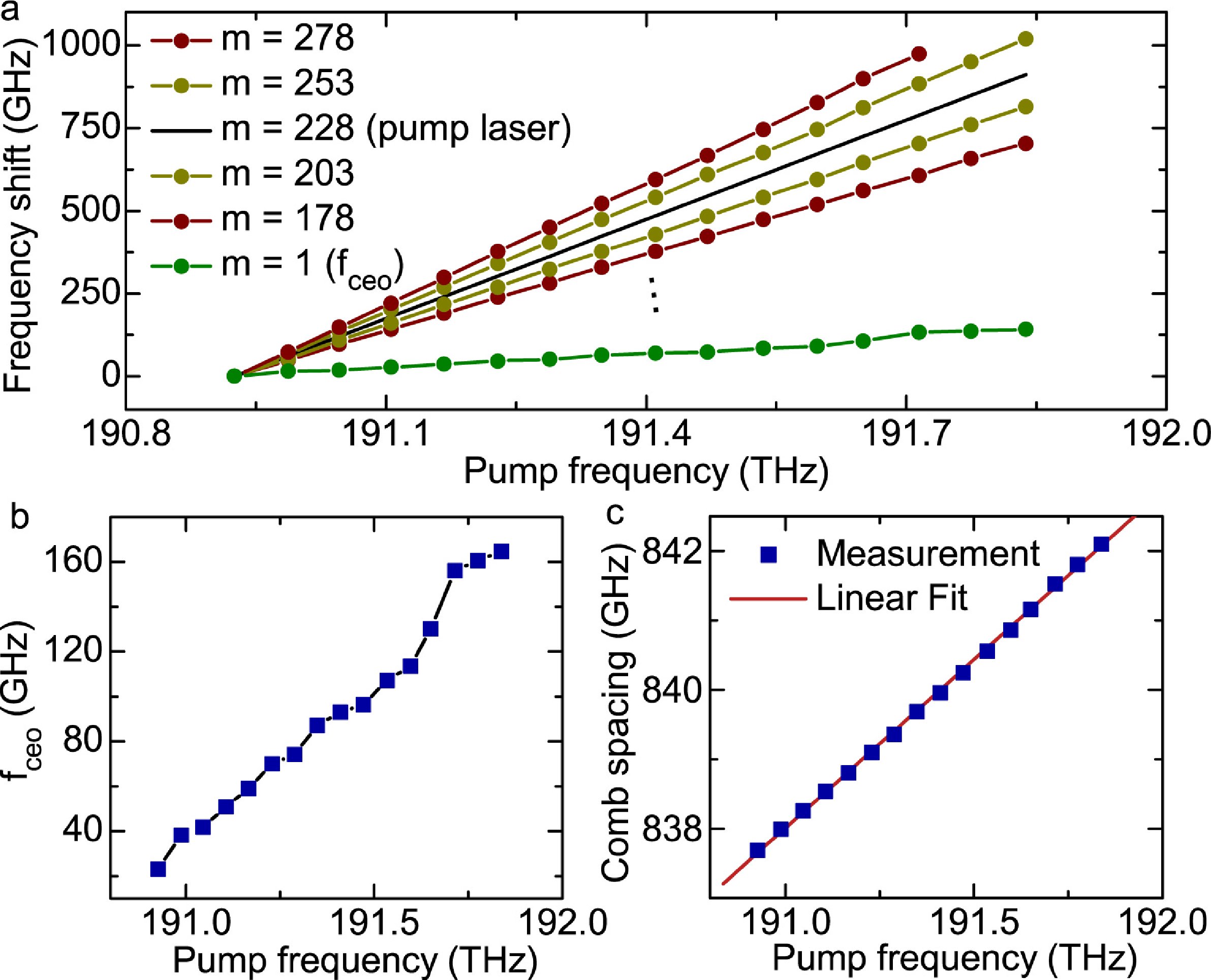}
\end{center}
\caption{\textbf{Microresonator based frequency comb tuning parameters.} \textbf{a.} Measured frequency shift of the $\pm25^{\mathrm{th}}$ and $\pm50^{\mathrm{th}}$ sideband with respect to the pump laser at different pump laser detuning. The black line has a slope of $1$ and depicts the shift of the pump laser itself. The green line shows the shift of the carrier envelope offset frequency $f_{\mathrm{ceo}}$. \textbf{b,c.} Effect of the pump frequency detuning on $f_{\mathrm{ceo}}$ and comb spacing.}
\label{fig3}
\end{figure}

The remaining parameters are the linear ( $n=1.44$ ) and nonlinear ($n_{\mathrm{2}}=2.2 \times 10^{-20} \frac{\mathrm{m}^{2}}{\mathrm{W}}$) refractive index of fused silica and the speed of light in vacuum $c$. Subtracting the Kerr contribution leads to a mode shift of $960$~GHz that is thermally induced. This thermal shift $\Delta \nu_{\mathrm{therm}}$ can be described by
\begin{equation*}
\frac{\Delta\nu_{\mathrm{therm}}}{\nu}=a\cdot\Delta T
\end{equation*}
, where $a=6\times10^{-6}$~K$^{-1}$is a combined constant describing both the thermal expansion and thermally induced refractive index change of fused silica. Inserting the frequency of the pump laser $\nu =192$~THz this equates to a remarkably high temperature change of $\Delta T \approx 833$~K. However, this temperature is still well below the annealing point at $1140^{\circ}$C and the softening point at $1665^{\circ}$C of fused silica. This thereby demonstrates the stability of the approach which even under such strong heating allows stable and long term optical frequency comb generation.

On first notice, the data in Figure~\ref{fig2} implies that the whole comb is shifted uniformly, which would correspond to a pure carrier envelope offset frequency change. However, a closer inspection of the comb modes reveals that the mode spacing is also affected by the pump laser frequency variation. The reason for this is the change in intracavity power due to the varying detuning between pump laser and microcavity mode\citep{Del'Haye2008}. Making use of the high number of comb modes, the comb spacing can statistically be determined much more precisely than expected from the resolution bandwidth of the optical spectrum analyzers, which was set to $25$~GHz. Figure~\ref{fig3}b and \ref{fig3}c show the variation of the mode spacing and the carrier envelope offset frequency as a function of the pump laser frequency (all the data has
been derived from the optical spectra of the frequency combs). It can be seen that the influence of the pump laser detuning on the mode spacing is rather small with $4.83$~GHz$/$THz compared to the impact on the carrier envelope offset frequency which equates to $153$~GHz$/$THz. Figure~\ref{fig3}a depicts the measured shift of the $\pm25^{\mathrm{th}}$ and  $\pm50^{\mathrm{th}}$ comb modes as a function of the pump frequency and is displayed together with the calculated shift carrier envelope offset frequency (being the imaginary first comb mode, cf. Figure~\ref{fig1}e). The zero position of the frequency shifts is set to the position of the cold microresonator mode that is used for comb generation.

\begin{figure}[ptbh]
\begin{center}
\includegraphics[width=1\linewidth]{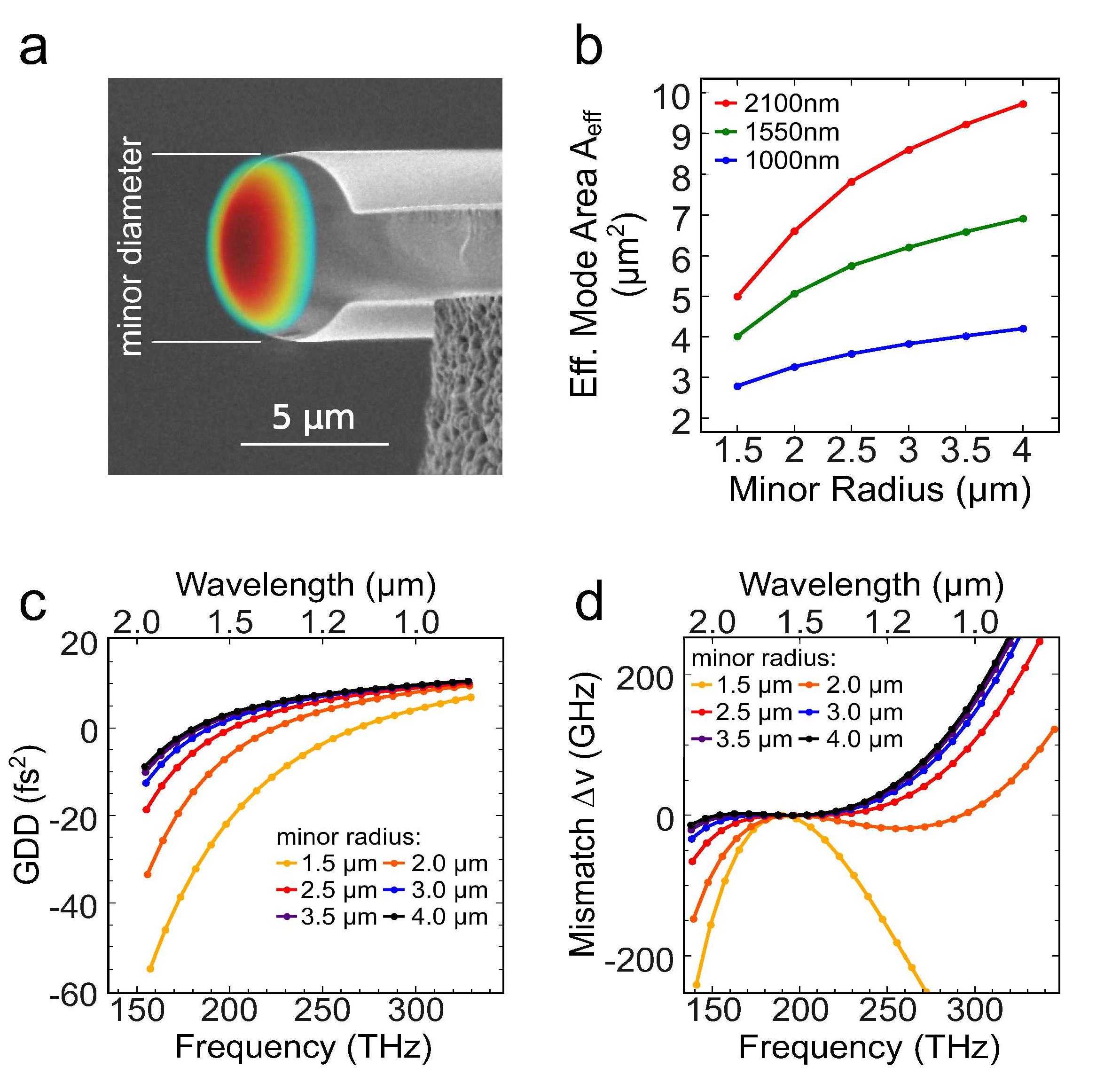}
\end{center}
\caption{\textbf{Simulation of mode volume and dispersion.} \textbf{a.} Simulated mode profile superimposed on a scanning electron microscope image of a cut microresonator. \textbf{b.} Simulated effective mode area for different wavelengths corresponding to the pump frequency and the high and low frequency ends of the experimentally generated octave spanning comb. Generally the effective mode area increases with the size of the microresonator's minor radius with longer wavelengths being affected the most. \textbf{c,d.} Simulated group delay dispersion (GDD) and mismatch $\Delta \nu$ between positions of equidistant comb modes and microresonator modes, subject to material and geometric dispersion within the resonator. Each evaluation point is marked by a dot and corresponds to a microresonantor mode. The simulated resonances are separated by ten free spectral ranges.}
\label{fig4}
\end{figure}

The geometry of a microresonator is critical for its potential for broad comb generation. Mode profile and resonance frequencies of different microresonator geometries were simulated using finite element methods\citep{Oxborrow2007}, taking geometrical and material dispersion into account. The geometries are defined by a fixed microresonator diameter of  $80$~$\mathrm{\mu m}$, determining the spacing of the comb modes, and a set of different minor radii, i.e. the crosssection radii of the silica toroid. Figure~\ref{fig4}a shows a simulated mode profile superimposed on a scanning electron microscope image of a microresonator, which has been cut in order to image the toroidal cross section. Based on these simulations the resonator's group delay dispersion (GDD) and the resulting wavelength dependent mismatch $\Delta\nu$ between positions of the equidistant comb modes and resonator modes (that are shifted from their equidistant positions due to dispersion) can be derived\citep{Del'Haye2009}.

Moreover, the effective mode area $A_{\mathrm{eff}}=\frac{\left(\int \vec{E}\cdot\vec{D}\,\mbox{d}A\right)^{2}}{\int \left(\vec{E}\cdot\vec{D}\right)^{2} \,\mbox{d}A}$, which is inversely proportional to the parametric gain bandwidth\citep{Stolen1982} can be calculated by integrating the numerical solution over the toroidal cross section plane (here, $\vec{E}$ is the electrical field and $\vec{D}$ the electric flux density). These results are depicted in Figure~\ref{fig4}b,c and d, respectively. The simulations show, that the extent of the minor radius - which is experimentally controllable - has considerable influence on effective mode area and dispersion of the resonator. With increasing minor radius the dispersion curve flattens, which is generally beneficial for broad comb generation as the absolute mismatch is smaller over a wider wavelength range, however, this has to be balanced with the increasing mode area, raising the power threshold\citep{Kippenberg2004a} for the four-wave
mixing process. The microresonator used for the measurements has a diameter of
$80$~$\mathrm{\mu m}$ and a minor radius of $2.8$~$\mathrm{\mu m}$, corresponding to an intermediate geometry, balancing limitations of dispersion and parametric gain.

In conclusion we have demonstrated octave spanning frequency comb generation directly in a monolithic microresonator for the first time. The presented comb is tunable over more than one mode spacing of the microresonator, which allows reaching every optical frequency in the wavelength range between $990$~nm and $2170$~nm with a comb mode. This work demonstrates that microresonator based frequency combs are promising for multi-channel telecommunication, metrology and various other applications such as spectrometer calibration and gas sensing. Especially for telecommunication and gas sensing, the high power per comb line of more than $100$~$\mu$W over a bandwidth exceeding $140$~THz is a great advantage compared to existing frequency combs. The intriguing ability to generate a full octave spanning frequency comb within a monolithic, low cost microresonator is a great step towards direct integration of optical frequency standards into compact microphotonic devices that may permit to bring an radio-frequency to optical link into a chip scale format, enabling entirely novel on-chip photonic functionality, unifying the fields of optical frequency metrology and nanophotonics. Moreover, the presented source has applications in optical coherence tomography, which require a broad bandwidth.

\section*{Acknowledgements}
T.J.K. acknowledges support from an Independent Max Planck Junior Research Group. This work was funded as part of a Marie Curie Excellence Grant (RG-UHQ) and the DFG funded Nanosystems Initiative Munich (NIM). We acknowledge the Max Planck Institute of Quantum Optics and P. Gruss for continued support. The microresonator samples were fabricated in the CMI cleanroom of the EPFL, Switzerland.

\bibliographystyle{naturemag}
\bibliography{references}

\end{document}